# Emerging Challenges in Molecular Paleontology: Misapplication of Environmental DNA Fragments and Misconception of "Deamination" as a Key Criterion for *In Situ* DNA Identification

Wan-Qian Zhao[1#], Shu-Jie Zhang[1], Zhan-Yong Guo[2], Zeng-Yuan Tian[3], Gang-Qiang Cao[3], Mei-Jun Li[4], Li-You Qiu[5#]，Jin-Yu Yang[6]，Yong-Kai Wang[1], Shu-Hui Zhang[1], Zhi-Fang Zheng[1], Min-Zhi Wu[1]

[#] Corresponding authors. Email: wqzhao@zzu.edu.cn; qliyou@henau.edu.cn

**Affiliations:**

[1] School of Life Sciences, Zhengzhou University, Zhengzhou, China

[2] College of Agronomy, Henan Agricultural University, Zhengzhou, China

[3] School of Agricultural Sciences, Zhengzhou University, Zhengzhou, China

[4] National Key Laboratory of Petroleum Resources and Engineering, College of Geosciences, China University of Petroleum (Beijing), Beijing, China

[5] College of Life Sciences, Henan Agricultural University, Zhengzhou, China

[6] Haining BoShang Biotechnology Co., Ltd, Haining, China

Abstract: This article critically examines the methodologies applied in ancient DNA (aDNA) research, particularly those developed by Dr. Pääbo's team, which have significantly influenced the field. The focus is on the challenges of distinguishing original *in situ* DNA (oriDNA) from environmental DNA (eDNA) contamination in fossil samples. Recent analyses indicate that even with rigorous extraction and sequencing protocols, a considerable amount of eDNA remains present, often misinterpreted as oriDNA. This misidentification risks the accuracy of species ascription and evolutionary interpretations derived from fossil analyses. The paper explores fossil preservation's physical and chemical dynamics, which allow eDNA from similar and disparate species to infiltrate bone matrices. We propose enhancements to methodological frameworks, such as broader BLAST database usage and stringent E-value criteria, to improve species-specific aDNA identification. Additionally, the article critiques the reliance on deamination patterns as a definitive marker for aDNA, suggesting a reevaluation of this criterion due to its inconsistency and the potential for misleading sequencing results. Ultimately, our findings advocate for a more cautious and refined approach to aDNA research, ensuring more reliable and verifiable scientific outcomes.

Keywords: ancient DNA, original *in situ* DNA, environmental DNA, deamination

Since 1985, Dr. Pääbo has undertaken many genetic studies focusing on early extinct humans and various ancient populations. Central to Pääbo's research methodology is that a fossil holds DNA representative of a single individual, consistent with the fossil's species classification – original, *in situ* DNA (oriDNA). The fossils, however, are often compromised by environmental DNA (eDNA) from microorganisms, while DNA contamination from modern humans arises predominantly during the excavation and processing of fossils. Implementing stringent anti-pollution procedures can effectively prevent such contamination during these operations. *Ex situ* DNA can be excluded by selecting DNA fragments that exhibit minimal natural damage (e.g., "deamination"). In 2013, Pääbo's team created a comprehensive method for ancient DNA (aDNA) research that built upon prior work, encompassing steps such as DNA extraction, probe enrichment of fragments, screening of fragments, sequence alignment, and genome assembly (1, 2). This methodological framework has led to progressive research results published in numerous prestigious journals (3-10).

In recent years, significant issues have been revealed with the methodology used in this field, raising important questions. We will carefully analyze the underlying assumptions and each step of Pääbo's method and share our insights.

**Fossils Contain Many DNA Fragments from *Ex Situ* Species**

In 2013, Pääbo's team established a foundational research method and applied it to the study of the "Tianyuan Cave Man." They operated under the premise that the aDNA extracted from the bone fossils was exclusively derived from the bones of a single individual within the species *Homo sapiens*. However, it was ultimately discovered that the proportion of endogenous skeletal DNA did not exceed 0.03%, and no detailed analysis of exogenous DNA was conducted (1). In the same year, significant amounts of human mitochondrial DNA (mtDNA) were also identified in ancient cave bear bone samples (2). This raises several questions: Given the rigorous aDNA protocols employed by Pääbo's team, how is it that there is a notable presence of non-human DNA in the extracted human bone samples, and how is human DNA present in animal bones? Are these DNA fragments signs of contamination from outside sources after the fossils were excavated? Or did the DNA fragments exist before the fossils were unearthed, originating from other individuals of the same or different species and entering the internal space of the fossils over various periods?

Pääbo's team overlooked two critical factors. Firstly, fossils are primarily composed of rock, which can feature various physical spaces, such as small holes, tunnels, and cavities that interconnect with the external environment by gaps. Mammalian bones contain cavernous sinus bundles composed of trabeculae with spaces in between. As a result, these natural internal spaces can allow foreign DNA to enter fossilized bones. Recently, we developed a calculation method to ascertain the internal space volume ratio of the Lycoptera fossil (11), a key representative of the Jehol Biota.

Secondly, due to the processes of diffusion and osmosis, when the external environment becomes dry, various molecules not embedded within the fossil can be lost as internal liquid water drains away. Subsequently, after water flows back into the environment, foreign molecules may be carried into the dry gaps within the fossil and become retained. Additionally, some molecules might be encapsulated by minerals in the water and deposited onto the inner surfaces of the rock. However, when the environment dries out again, free molecules may escape. This cycle of drought and rain will persist until gaps are filled. A thought experiment indicates that eDNA will prevail among these molecules. Consequently, there is a continuous interchange between older and newer eDNA within the fossil, resulting in a phenomenon, named "less old and more new molecules" (11).

eDNA predominantly originates from the dominant species in a given habitat. This creates a scenario where, in areas permanently inhabited by advanced primates, and especially by *Homo species*, due to their social practices of burying the remains of deceased companions, these long-term and ongoing burial activities continuously contribute their DNA to the environment, along with DNA fragments from associated species, such as animals obtained through hunting or farming.

Fossils are a type of rock that naturally contains internal space. As one observes the rain this morning nourishing the Cretaceous rocks, it's important to recognize that rainwater flow can introduce contemporary eDNA into ancient fossils. Similarly, precipitation from the past had a comparable effect. Consequently, Cretaceous fossils may contain eDNA from later periods, such as the Cenozoic era, and this phenomenon should not be a source of confusion.

We aligned DNA fragments extracted from Lycoptera fossils dating back 120 million years ago with all known sequences (version 5 of the nucleotide sequence database) on our local computer station. For each BLAST result, we selected the top hit based on the E-value rank (the smallest one), identifying over 230,000 DNA fragments that matched the genomes of Hominoidea (apes). Additionally, we acquired 266 fragments that aligned with the Pan-genomes through the "minimal E-value mode" during our online search. Additionally, using the 'MS mode' online, the results indicated that 22 fragments uniquely aligned with the Pan-genome. This means that the first matched species, *Pan troglodytes*, exhibits an E-value difference greater than 1E-02 compared to the second matched species, a non-Pan species (see the Methods in reference 11). However, these DNA fragments do not possess 100% similarities with the Pan-genomes, leading to notable discrepancies. For example, the Identity or Cover of these "difference" fragments can fall below 90% of the modern Pan sequence. Notably, there is no detectable "deamination" (Table S3B of reference 11; full sequencing data available upon request). Therefore, the result indicates that these fragments do not originate from modern Pan-genomes but rather from the genomes of early Homininae species (Figure 1A and 1B).

The cost of artificially synthesizing and faking these "difference" fragments could be as high as millions of dollars, making such a malicious attempt prohibitively expensive (11, 12).

In summary, we can draw the following conclusions: Given that there is no existing population of chimpanzees and the fossil evidence from the Jehol area, these DNA fragments must originate from ancient Homo species or close relatives, having gradually penetrated into the specimens over the years since the fossil's formation. The "deamination" damage is not a prerequisite for assessing aDNA fragments. Hence, we contend that the assembled genome of the so-called "Tianyuan Cave Man" is likely derived from unclassified Homo species from various periods, including modern humans, or possibly a mix of genomes from other genera. Additionally, the findings of nearly all studies employing Pääbo's method warrant a reevaluation to move beyond this challenging situation.

**Probe Enrichment Does Not Exclude Sequences from *Ex Situ* Species**

According to the method developed by Pääbo's team, the hybridization temperature for the enrichment probe and the target fragment is set between 62-65°C (1, 13). However, since the average length of the target fragments in this experiment exceeds 30 bp, the likelihood of DNA double-stranded base mismatches increases as the fragment length increases during the hybridization process. This poses challenges for the enriched probe in achieving species-specific capture, raising the possibility that the extracted DNA may contain gene fragments from non-target species in the environment. The author did not conduct any corresponding exclusion tests to address this concern.

The authors conducted a BLAST of the captured DNA fragments exclusively against Homo mitogenomes, thereby categorizing these fragments as mtDNA sequences. Such alignments, which rely solely on mitogenomes as the reference, may inadvertently include eDNA fragments that share sequence similarities with them. Furthermore, the authors did not adequately highlight the significance of the E-value in selecting BLAST results. A high E-value in BLAST results indicates that the match may fall outside a credible range.

We found only one aDNA sequence (SI Fig S3A: CCTATAGCACCCCCTC) in the referenced papers of Pääbo's team ([1](#)). Upon comparing this sequence with all known genomes (version 5, nucleotide sequence database), we observed that it matches 100% with genome sequences from various animals, plants, and bacteria ([Figure 2](#)). It is well established that bacterial genomes are the primary contributors to environmental eDNA in soil. Consequently, the mitogenome (KC417443) of "Tianyuan Cave Man" may be mixed with gene fragments from many other species. Additionally, the extensive length of chromosome 21, which spans 48.5 Mb, may result in more fragments displaying higher similarity to DNA sequences from various animals, plants, and microorganisms. Therefore, the assembled ERP002037 could face significant issues with mixed DNA fragments from multiple species.

DNA extracted from fossils often contains significant eDNA contamination, leading to the capture of eDNA of non-target species. Modern human sequences, especially mtDNA sequences, are highly similar to those of ancient humans and various mammals, increasing the risk of misusing eDNA sequences.

To enhance the accuracy of sequence identification, it is essential to trace and evaluate DNA fragments before splicing. Specifically, all available genomic data should be used as a reference, and the fragments that lack "uniqueness" in comparison should be discarded. After this screening, the remains can be spliced or partially spliced, followed by phylogenetic testing to ensure the reliability of the results.

## "Deamination" Should Not Be Viewed as an Essential Criterion for Identifying aDNA

The extent of DNA fragment "deamination" is closely associated with the chemical composition of fossils (rocks) and their geological history. Lindahl suggested that low temperatures, rapid drying, and high salt concentrations can significantly reduce the degradation of aDNA, highlighting that the "deamination" process occurs exclusively in the presence of liquid water ([14](#)). This finding underscores the necessity of liquid water as a crucial condition for aDNA "deamination". Furthermore, DNA "deamination" is affected by many factors within the fossil and surrounding environment, making it a non-uniform process.

When studying aDNA, one must consider the physical and chemical properties of the fossil and the surrounding environmental conditions, as well as the integrity of the aDNA fragments in terms of protection and sealing. If the spaces between DNA molecules are filled and sealed with mineral salts - common constituents of certain sedimentary rocks - this can effectively inhibit "deamination" reactions. It is essential that various chemical groups within the molecule, including the bases and glycosides, remain shielded from exposure to liquid water, as this is vital for preventing degradation. For example, many volcanic sedimentary rock fossils exhibit well-preserved nuclear structures, suggesting that oriDNA remains (15-17). This preservation is most likely due to rapid burial and dehydration around volcanic ash during fossil formation. The Lycoptera fossils we investigated are classified as volcanic tuff fossils, serving as a pertinent example (11).

Pääbo's team focused exclusively on DNA fragments that were significant "deamination" as oriDNA (10). However, this screening approach presents three significant and unresolved challenges. First, it is unclear how to accurately determine the age of aDNA based on the extent of "deamination" in the fragments. Additionally, distinguishing a specific DNA sequence as belonging to Neanderthals from 50,000 years ago versus modern humans from a more recent period, such as 10,000 years ago, poses a significant problem. Therefore, relying solely on "deamination" characteristics of DNA is inadequate for ruling out modern human contributions. Second, we identified numerous aDNA fragments from ancient ages that did not display discernible "deamination" reactions (Table S3A and S3B of reference 11). This observation suggests that "deamination" should not be deemed a necessary criterion for identifying aDNA. Third, when a fragment exhibits considerable "deamination," the original genetic information becomes significantly distorted. Consequently, the E-value for the corresponding BLAST hit often becomes elevated, exceeding the confidence thresholds set by BLAST. For these reasons, selecting such fragments for further research is generally inadvisable. Lastly, the chosen fragments feature short effective sequence lengths for BLAST that can simultaneously match the genomes of many species, including the target species. Given that the DNA library mainly consists of non-*in situ* fragments, including DNA from lower organisms, the likelihood of eDNA fragments aligning with the target genome is significantly increased. This creates a risk of misuse, highlighting the vulnerabilities associated with this method.

In summary, Pääbo's approach suffers from four important shortcomings: First, it fails to account for the possibility that eDNA may have continuously infiltrated bone fossils since their formation. Consequently, the oriDNA referenced in their papers cannot eliminate the risk of contamination from DNA belonging to the same or different species across various periods. Second, relying exclusively on the genome of the target species as a comparative reference leads to the misclassification of numerous non-target DNA fragments as oriDNA. Third, the team overlooks that BLAST results with higher E-values are outside the credible interval, which results in the inappropriate use of non-*in situ* fragments. Fourth, using "deamination" as a primary criterion for identifying aDNA is fundamentally flawed. This approach not only inflates the overall E-value and narrows the confidence interval of the results by favoring non-*in situ* fragments but also selectively excludes long strands of oriDNA that contain valuable genetic information.

**Exploring aDNA Research Methods**

In our view, meticulous tracking of the "unique" species origin and taxonomic classification of each DNA sequence is essential in aDNA research to differentiate oriDNA from the DNA of non-target species that may be present (11, 12). To begin with, the reference genome used for BLAST alignment should include all known genomes (referred to as Version 5 of the Nucleotide Sequence Database), rather than being limited to the target species. The BLAST results should indicate an E-value for the target species with less than 1E-07 to reduce the risk of systematic identification errors. In addition, if the BLAST hits multiple species, the target one must be ranked highest and have an E-value that differs from the E-value of the second match by more than 1E-02; otherwise, the identification should be deemed inconclusive. These measures are vital for confirming the "uniqueness" of the species' origin and are crucial for the selection of oriDNA.

Furthermore, the "deamination" of bases within the DNA strand can lead to distortions in genetic information. Severely deaminated fragments often result in a high mismatch between query and subject sequence, producing elevated E-values in comparison results, which can be misleading. The phenomenon of "deamination" is more prevalent at the 3' end of the DNA chain, while it appears sporadically in the middle, interspersed among normal bases (1). This variability makes it challenging to determine whether the adenine (A) and thymine (T) in a fragment are the product of "deamination" or point mutations. Notably, several key studies employing Pääbo's method have reported fragment lengths that are consistently too short (2, 3), which inevitably leads to multiple matches with environmental species in the BLAST hits, thus failing to meet a proper E-value and complicating the confirmation of the DNA sequence's uniqueness. For instance, in the four samples from the "Tianyuan Cave Man," the average fragment lengths were only 64 bp, 52 bp, 59 bp, and 42 bp (1). These short fragments showed significant "deamination". The sequences suitable for BLAST became shorter and of poor quality, leading to high E-values that lacked credibility. This necessitates a more thorough analysis and re-evaluation. Therefore, we recommend avoiding those fragments with pronounced "deamination" for research use.

In summary, the main flaw in Pääbo's method is its inability to acknowledge that water, as the primary medium, can replace the internal DNA of fossils with external eDNA. Furthermore, it does not realize the influence of external eDNA fragments that may be closely similar to the DNA sequences of *in situ* species, significantly affecting the results. The assumption that aDNA fragments will inevitably be "deamination" reflects an overly expected perspective.

Despite the unchanged relative fossil morphology of the species, its internal DNA origins have shifted. While the fossil retains some oriDNA, most of the internal space may have been occupied by eDNA molecules from later periods, including fragments of aDNA from various historical epochs. This scenario undoubtedly complicates research in molecular paleontology. However, the presence of aDNA fragments in fossils and sedimentary rocks also has beneficial consequences, greatly expanding the possibilities of paleontological molecular biology. For example, collecting aDNA from sedimentary sand and gravel surrounding fossil remains presents a promising approach (18).

We propose implementing a trinity approach to the search for aDNA not just in fossils, but also in rocks, particularly source rocks with internal space, and in petroleum from the former habitats of specific ancient species. Preliminary aDNA data analysis of rock formations can assist in searching physical fossils, especially those with difficult-to-recognize shapes. This approach significantly broadens the scope of molecular paleontology research.

To uphold the rigor of scientific research and facilitate the re-evaluation of studies utilizing Pääbo's method, we recommend that Pääbo's team promptly share DNA sequences and sequencing data associated with the data in their publications. This will provide crucial materials for comparative analyses and enable accurate identification of oriDNA, along with the necessary technical support.

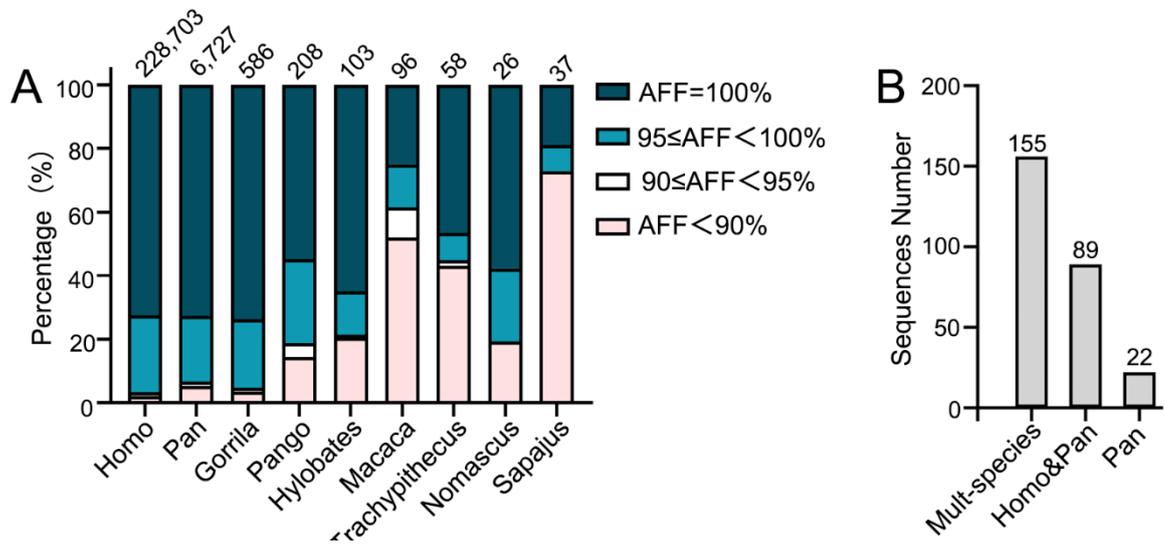

**Figure 1. Analysis of 236,544 Sequences Aligned to the Hominidae Family (Apes)**

These sequences were selected from the sequencing data using the "minimum E-value mode". **A**: Distribution of sequence affinities (Query) among Hominoidea species, calculated as Affinity = Cover × Per Ident × 100. **B**: The 266 sequences were extracted from 6,727 aligned sequences corresponding to the Pan-genome (with 38% ≤ Affinity ≤ 90) and subsequently screened and classified using the "MS Model" (11, 12). The results show that 155 sequences (IDs: 379-533) map to the genomes of multiple Hominoidea species, 89 sequences (IDs: 290-378) align with the genomes of humans and chimpanzees, and 22 sequences (IDs: 534-555) uniquely match the Pan-genomes. These sequences can be found in Table S3B of reference 11.

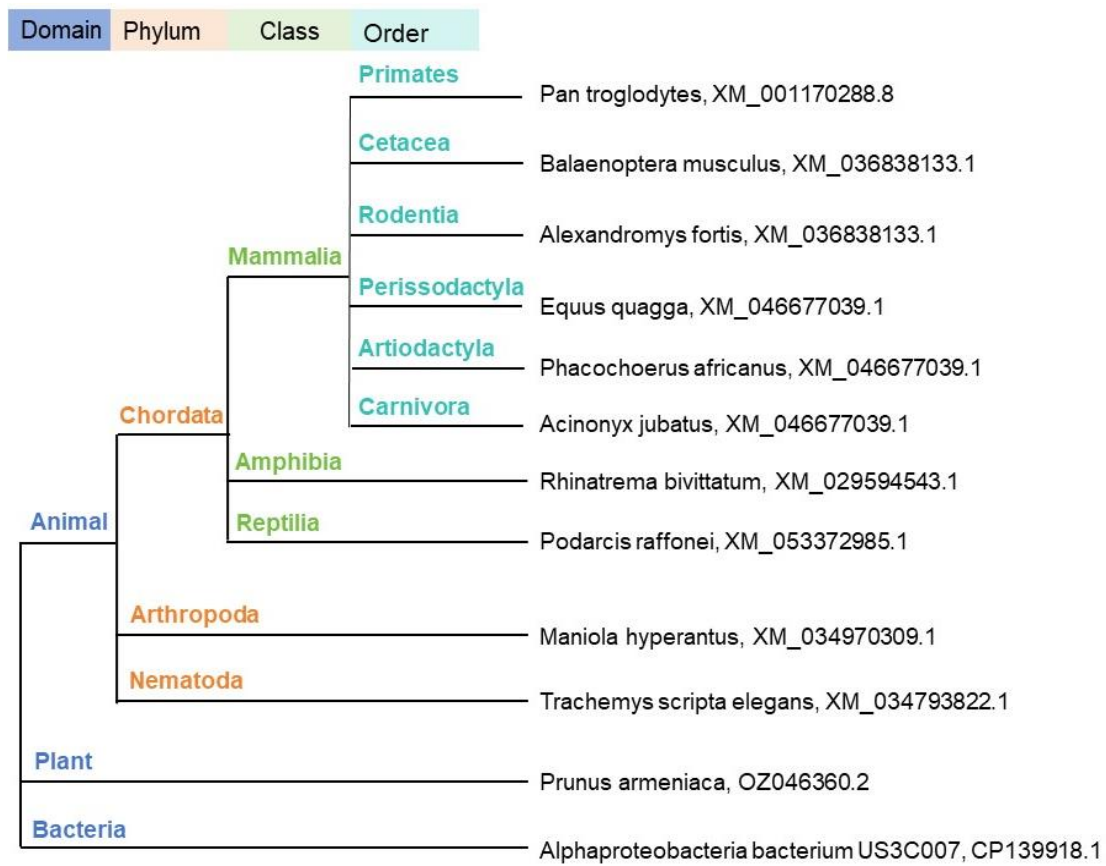

**Figure 2. Results of Nucleotide BLAST Analysis**

The figure presents the results of a nucleotide BLAST analysis, highlighting multiple matches for the sequence from SI Fig S3A: CCTATAGCACCCCCTC reported by Pääbo's team (1). When we compared this sequence to all known genomes available on the NCBI database, we discovered that it aligns perfectly, showing a 100% match with genome sequences across various domains of life.

180 Million Years of Genomic Stasis in Royal Ferns. *Science* **343**, 1376-1377 (2014).
18. D. Zhang *et al.*, Denisovan DNA in Late Pleistocene sediments from Baishiya Karst Cave on the Tibetan Plateau *Science* **370**, 584-587 (2020).